\documentclass[10pt,dvips,twoside]{article}
\usepackage{amstext,amsmath,amssymb,amsfonts}
\usepackage{cite}
\usepackage{graphicx}
\usepackage{epsfig}
\usepackage[dvips,usenames]{color}
\usepackage{color}
\usepackage{rotating}

\usepackage[all]{xy}
\usepackage{tabularx}
\usepackage{fancyhdr}

\usepackage{latexsym}

\newcommand{\ba}{\begin{array}}
\newcommand{\ea}{\end{array}}

\newtheorem{theorem}{Theorem}[section]
\newtheorem{coroll}[theorem]{Corollaire}
\newtheorem{lemma}[theorem]{Lemma}
\newtheorem{prop}[theorem]{Proposition}
\newtheorem{fit}[theorem]{Definition}

\newcommand{\betheo}{\begin{theorem}}
\newcommand{\entheo}{\end{theorem}}

\newcommand{\becor}{\begin{coroll}}
\newcommand{\encor}{\end{coroll}}

\newcommand{\belem}{\begin{lemma}}
\newcommand{\enlem}{\end{lemma}}

\newcommand{\befit}{\begin{fit}}
\newcommand{\enfit}{\end{fit}}

\newcommand{\beprop}{\begin{prop}}
\newcommand{\enprop}{\end{prop}}

\def \k0 {\frac{1 }{4 \pi \epsilon_0}}

\def \bc {\begin{center}}
\def \ec {\end{center}}

\def\half{\mbox{\small{$\frac{1}{2}$}}}

\def \pt {\ .}

\newcommand{\virg}{{\textrm{\ ,}}}

\def \dis {\displaystyle}

\def \id {\hbox{\bf \large 1} \!\! \hbox{\bf \large \sf I}}

\def \IR {\rm I \! R}
\def \IC {{\sf I} \!\!\! {\rm C}}

\def  \ni {\noindent}

\def  \lv {\left|}
\def  \rv {\right|}

\def  \li {\left<}
\def  \rs {\right>}

\newtheorem{theo}{Theorem}[section]

\newtheorem{definition}[theo]{Definition}

\newtheorem{example}[theo]{Example}
 
 \numberwithin{equation}{section}

\newtheorem{remark}[theo]{Remark}

\newcommand{\cqfd}
{%
\mbox{} \nolinebreak \hfill \rule{2mm}{2mm} \medbreak
\par
}

\newcommand{\bedefi}{\begin{definition}$\!\!\!$ \rm }
\newcommand{\findefi}{ \end{definition}}

\newcommand{\beex}{\begin{example}$\!\!\!$ \rm }
\newcommand{\enex}{ \end{example}}

\newcommand{\berem}{\begin{remark}$\!\!\!$ \rm }
\newcommand{\enrem}{ \end{remark}}

\newcommand{\be}{\begin{equation}}
\newcommand{\en}{\end{equation}}

\newcommand{\bea}{\begin{eqnarray}}
\newcommand{\ena}{\end{eqnarray}}

\newcommand{\beano}{\begin{eqnarray*}}
\newcommand{\enano}{\end{eqnarray*}}

\newcommand{\bee}{\begin{enumerate}}
\newcommand{\ene}{\end{enumerate}}

\newcommand{\bei}{\begin{itemize}}
\newcommand{\eni}{\end{itemize}}

\newcommand{\betab}{\begin{tabular}}
\newcommand{\entab}{\end{tabular}}

\def\dag{\dagger}

\begin{document}

 \setcounter{section}{0}
\setcounter{equation}{0}
\setcounter{figure}{0}
\setcounter{table}{0}
\setcounter{footnote}{0}


\vspace*{10mm}

\begin{center}
{\bf\Large  Shape invariant potential formalism for photon-added coherent state construction
} 

\vspace{10pt}
\end{center}



\begin{center}
\vspace*{10pt}

{\large\sc Komi Sodoga  $\!^{\rm a,b}$, Isiaka Aremua  $\!^{\rm a,b}$ and
   Mahouton Norbert Hounkonnou  $\!^{\rm b}$}
\\[3mm]
$^{\rm a}$ \textsl{\small  Universit\'e de Lom\'e,  Facult\'e des Sciences, D\'epartement de Physique,\\
 Laboratoire de Physique des Mat\'eriaux et de M\'ecanique Appliqu\'ee,
 02 BP 1515 Lom\'e, Togo  \\
\hspace*{3mm}E-mail: ksodoga@univ-lome.tg, iaremua@univ-lome.tg}
\\[2mm]
$^{\rm b}$ \textsl{\small University of Abomey-Calavi, International Chair in Mathematical Physics   and Applications (ICMPA), 072 B.P. 050  Cotonou, Benin \\
\hspace*{3mm}E-mail : norbert.hounkonnou@cipma.uac.bj
}
\end{center}


\vspace{15pt}

\begin{quote}
An algebro-operator approach,  called  shape invariant potential  method,  of constructing  generalized coherent states for photon-added  particle system is presented. Illustration is given on P\"oschl-Teller potential.
\end{quote}

\thispagestyle{plain}
\fancyfoot{}


\renewcommand\headrulewidth{0.5pt}
 
\section*{Introduction}
\label{sect_intro}

 Coherent states (CS) play an important role in many fields of quantum mechanics since his early days. 
These states  were first introduced by Schr\"odinger \cite{Schrodinger} since 1926 for  the harmonic oscillator. Then followed  decades of intensive works in order to extend the CS concept to other types of exactly solvable systems  \cite{Sodoga, Aremua1, Aremua2, Glauber, Pere}. 
 It was shown in 1980's 
that a large class of these solvable potentials are characterized by a single  property,
i.e., a discrete reparametrization invariance, called shape-invariance \cite{Gend, Dab, Dutt5, Suk}, introduced in the framework  
of the supersymmetric quantum mechanics (SUSY QM) \cite{Infeld, Cooper}. It was
then shown that shape invariant potentials (SIP) \cite{Dab, Suk}  have an underlying algebraic structure and
the associated Lie algebras were identified\cite{Bal, Alex}. Using this algebraic structure, a general
definition of coherent states for shape invariant potentials were introduced by different
authors \cite{Fukui, Alex}.

 In 1980's, a  new class of nonclassical states, known as photon-added coherent states (PACS), was introduced by Agarwal and Tara \cite{Aga}. These states which are intermediate states between CS and Fock states 
are constructed by repeated application of the creation operator on an ordinary CS. The PACS have known
 a great interest, as shown the  different extensions \cite{Penson, 6derniers,Popov, popov01, Kinani} and  applications  of the concept in various field of physics \cite{Dakna, Ban}. 

In a recent work \cite{SAH_prep}, we  constructed  photon-added CS for  SIP and investigated different cases following the Infeld-Hull \cite{Infeld} classification. 

In this contribution paper, we aim at providing a rigorous mathematical formulation of the CS and their photon-added counterparts for SIP. We  apply this formalism to P\"oschl-Teller potentials of great importance in atomic physics.  The diagonal $P$-representation of the density operator
$\rho$ is elaborated with  thermal expectation values. This computation gives value on  the use of
 Meijer G-functions. Novel results are obtained and discussed.

The paper is organized as follows. In section 2, we review the concepts of SUSY QM factorization, give the algebraic formulation of shape invariance condition, and define the generalized  shape-invariant potential coherent states (SIPCS). In section 3, we construct the photon-added shape-invariant potentials coherent states (PA-SIPCS) by successive applications of the raising operator on the SIP-CS.  We calculate the inner product of two different PA-SIPCS in order to show that the obtained states are not mutually orthogonal.  In contrast, we prove that these states are normalized.   The resolution of unity is checked.   Finally, we study the  thermal statistical properties of the PA-SIPCS in terms 
  of  the  Mandel’s Q-parameter. In section 4,  P\"oschl-Teller potentials are investigated as illustration. We end, in section 5, with some concluding remarks.

\section{Mathematical formulation of  SUSYQM: integrability condition and  coherent state construction}
\ni In this section,  we introduce  the SUSY QM factorization method \cite{HSA} (and references therein), give the integrability condition, known as  shape invariance condition, and define the associated generalized  coherent states.

Let 
   ${\cal{H}} = L^2(]a,  b[, \  dx)$ be the  Hilbert space  with the
inner product defined by :
\bea \label{g1}
 \langle u, v \rangle := \int_a^b {\bar u}(x)v(x)dx, \quad
  \forall\, \, u, v \in {\cal H},
\ena
where ${\bar u}$ is the complex conjugate of $u$.
Consider  on ${\cal{H}}$ the  one-dimensional  bound-state Hamiltonian ($\hbar = 2m = 1 $)
\bea \label{g2}
{H} = -{d\over dx}^2 + {V}(x), \quad x \in ]a, b[ \subset \IR
\ena
 with the domain
\bea \label{g3}
 {\cal D}({H}) = \left\{ u \in {\cal H}, \quad  - u''+ V  u
 \in {\cal H} \right\},
\ena
where $V(x)$ is a real continuous  function on $]a, b[$.
Let us  denote $E_n$ and ${ \Psi}_n$  the eigenvalues and eigenfunctions of ${H}$,
 respectively.
 Let  the first-order differential operator $A$ be  defined by:
\bea \label{g4}
{ A = {d\over dx} + W(x), \  \textrm{with the domain}\ 
{\cal D}(A) = \left\{ u \in {\cal H}, \quad   u'+  W u
 \in {\cal H} \right\},}
\ena
 ${W}(x)= -\dis {d\over dx}[\ln({\Psi}_0)] $ is a  real continuous  functions on $]a, b[$. 
The adjoint operator  $A^\dag$  of  $A$ is defined on   \cite{Teschl}:
\bea \label{g5}
  {\cal D}(A^\dag) = \left\{ u \in\, {\cal H}|\  \exists \,
  \tilde v \in \, {\cal H}:
  \langle Au, v\rangle = \langle u, \tilde v \rangle \quad  \forall
   u \, \in {\cal D}(A)\right\}, \quad  A^\dag v = \tilde v .
\ena
We infer ${\cal D}(A)$ dense  in  ${\cal H}$  since
$H^{1,2}\left(]a, b[, \, \rho(x) d x\right)$
 is dense in  $ {\cal H} $ and
$H^{1,2}\left(]a, b[, \, \rho(x) d x\right) \subset {\cal D}(H) $, where
  $H^{m,n}(\Omega)$ is the   Sobolev spaces of indices $(m, n)$.
We  assume that the  operator $A$  is  closed in ${\cal H}$.
The explicit expression of $A^\dag$ is given through the following theorem.
\betheo
Suppose  the following boundary condition:
  \bea \label{g6}
   u(x)\, v(x)\Big|_a^b
  = 0, \quad \forall \, \, u \, \in\, \,  {\cal D}(A) \, \,\textrm{and} \, \,
   v \, \,
     \in {\cal D}(A^\dag)  ,
  \ena
   is verified. Then the operator  $A^\dag$ can be written as
\bea \label{g7}
 A^\dag =  \left[-{d\over dx} + W(x) \right].
\ena
 \entheo
{\bf Proof:} The proof follows as 
 \beano
  \langle A{\bar u}, v \rangle & \equiv&
                                   \int_a^b \left[{\bar u}'(x) + W(x)   {\bar u}(x)\right]v(x) dx \\
 & = &     {\bar u}(x) v(x)\Big|_a^b + \int_a^b {\bar u}(x)\left[-{d\over dx} + W(x)\right] v(x) dx   \\
 & = &     \langle {\bar u} , A^\dag v \rangle \quad \textrm{ for any}\quad
 u\in  {\cal D}(A), \quad   v \in  {\cal D}(A^\dag). 
 \enano
\cqfd
\ni Let    $H_1$ and  $H_2$ be the product operators $ A^\dag A $ and
  $A \, A^\dag$, respectively, with  the corresponding domains
  \bea \label{g8}
  {\cal D}(H_1) & = & \left\{u \  \in  \  {\cal D}(A),\
  v = A u\   \in  \ {\cal D}(A^\dag)\ {\textrm  and}
   \  A^\dag v  \ \in  \  {\cal H} \right\},  \cr \label{g9}
  {\cal D}(H_2) & = & \left\{u \  \in  \  {\cal D}(A^\dag),\
  v = A^\dag u\   \in  \ {\cal D}(A)\ {\textrm  and}
   \  A v\ \in  \  {\cal H} \right\} .
  \ena
  Remark that  \[H^{1,2}\left(]a, b[, \, dx\right) \subset
    {\cal D}(A) \subset  {\cal D}(A^\dag) .\]
 Then
    \[{\cal D}(H_1), {\cal D}(H_2)
  \supset  H^{2,2}\left(]a, b[, \, dx\right).\]
   We infer then that ${\cal D}(H_1)$
   and ${\cal D}(H_2)$ are dense in $ {\cal H}$.
The following theorem gives additional  conditions  on $W$ so that the operator $H$ factorizes in terms  of
$A$ and $A^\dag$.
\betheo \label{theo2}
 Suppose that   the function $W$ verifies the Riccati type equation:
 \bea \label{g09}
   V - E_0 = W^2 - W'.
   \ena
 Then  the  operators
   $H_{1,2}$   are  self-adjoint, and:
 \bea   \label{g10}
 \begin{array}{l} H_1  =  A^\dag A = H - E_0 =   -\dis  {d^2 \over dx^2} +  W^2 - W',  \quad
  H_2  = A \, A^\dag = - \dis {d^2 \over dx^2} + W^2 + W'.
   \end{array}
   \ena
 \entheo
 {\bf Proof}\quad
The operators $A^\dag A$ and $A\, A^\dag$ are self-adjoint  since
$A$ and   $A^\dag$ are mutually  adjoint     and $A$ is closed  with
${\mathcal D}(A)$ dense in ${\cal H}$.
 From the definitions (\ref{g4}) et (\ref{g7}) of the differential operators $A$ and $A^\dag$, we have the following products
 \beano
 \, A^\dag\, A = - \dis {d^2 \over dx^2}  + (W^2 - W'),  \quad 
 \, A\, A^\dag = - \dis {d^2 \over dx^2}  + (W^2 + W').
 \enano
 The equation (\ref{g09})  are  readily
 deduced from the above relations  and (\ref{g10}). \cqfd
\ni We can rewrite  the operators $H_{1,2}$ as:
 \bea \label{g11}
 H_{1,2} = -  {d^2 \over dx^2}  + V_{1,2} ,\quad
\textrm{where} \quad V_{1,2} =  W^2 \mp W'.
 \ena
 In SUSY QM terminology, $H_{1,2}$ are called SUSY partner Hamiltonians;  ${V}_{1,2}$ are  called SUSY partner potentials,  and  the function $W$ is called the
 superpotential.\\\\
Now  let us establish some results showing that the eigenvalues of partner Hamiltonians are positive definite $(E_n^{1,2} \ge 0)$ and isospectral, i.e, they have almost the same energy eigenvalues, except for the ground state energy of $H_1$ \cite{Cooper}.
\beprop \label{prop1}
 The eigenvalues of  $H_1$ and  $H_2$ are non negative
 \beano E_n^{(1)} \ge 0,\quad  E_n^{(2)}\ge 0.\enano
\enprop
{\bf Proof}:  Let $E_n^{(1)}$  be an eigenvalue of $H_1$
corresponding to the eigenfunction $\Psi_n^{(1)}$.
 In Dirac
notation, this reads as   $H_1 | \Psi_n^{(1)} \rangle = E_n^{(1)}
|\Psi_n^{(1)} \rangle$. Then  $ \langle \Psi_n^{(1)}|A^\dag
A|\Psi_n^{(1)}\rangle
 = E_n^{(1)}\langle \Psi_n^{(1)}|\Psi_n^{(1)}\rangle$, i.e, $||A
 |\Psi_n^{(1)}\rangle||^2
= E_n^{(1)}|||\Psi_n^{(1)}\rangle||^2$. Therefore $E_n^{(1)} \ge 0$,
since $||A |\Psi_n^{(1)}\rangle||^2 \ge 0$ and
$|||\Psi_n^{(1)}\rangle||^2 \ge 0$. Similarly, one can show that
$E_n^{(2)} \ge 0.$ \cqfd
\beprop \label{prop2}
Let
$|\Psi_n^{(1)}\rangle$ and $|\Psi_n^{(2)}\rangle$ be the normalized
eigenstates of $H_1$ and $H_2$ associated to the eigenvalues
$E_n^{(1)}$ and  $E_n^{(2)}$, respectively.
 Then
 \beano
  A |\Psi_n^{(1)}\rangle = 0 \Longleftrightarrow E_n^{(1)} =0, \quad
   A^\dag |\Psi_n^{(2)}\rangle = 0 \Longleftrightarrow E_n^{(2)} =0 .
 \enano
\enprop
 {\bf Proof}:
 \beano
A |\Psi_n^{(1)}\rangle = 0 &\Longleftrightarrow & || A |\Psi_n^{(1)}\rangle||^2 = 0\\
 &\Longleftrightarrow &  \langle\Psi_n^{(1)} |A^\dag A| \Psi_n^{(1)}\rangle = 0\\
  &\Longleftrightarrow & E_n^{(1)} \langle\Psi_n^{(1)} | \Psi_n^{(1)}\rangle = 0\\
             &\Longleftrightarrow & E_n^{(1)} = 0.
             \enano
By analogy, one can show  that
 $  A^\dag |\Psi_n^{(2)}\rangle = 0 \Longleftrightarrow E_n^{(1)} =0$. \cqfd
\ni As a consequence of this proposition $E_0^{(1)} = 0$, since $A \Psi_0^{(1)} = A \Psi_0 =0$.
\beprop \label{prop3}
 If  $H_1$ admits a normalized eigenstate
$|\Psi_0^{(1)}\rangle$ so that  $E_0^{(1)} =0$,
 then  $H_2$ does not admit a normalized eigenstate
$|\Psi_0^{(2)}\rangle$ corresponding to the eigenvalue $E_0^{(2)} =
0$.
 \enprop
 {\bf Proof}: If $E_0^{(1)} =0$, then from the
proposition  \ref{prop2}, \    $ A \Psi_0^{(1)} =0$. We deduce from
this that
 \bea \label{g12} 
A A^\dag A \Psi_0^{(1)} = H_2 (A\Psi_0^{(1)}) = 0.
\ena
 Suppose that there exists a normalizable
eigenstate
 $|\Psi_0^{(2)}\rangle $ of  $H_2$ corresponding to  $E_0^{(2)} = 0$.
  It follows from  (\ref{g12})
 that  $|\Psi_0^{(2)}\rangle \propto A \Psi_0^{(1)} = 0$,
 that is inconsistent. \cqfd
 \ni This proposition shows that $H_2$ cannot possess a normalized
 state $\Psi_0^{(2)}$ corresponding to the eigenvalues $E_0^{(2)}=0$,
 since $E_0^{(1)} =0$, that means $E_0^{(2)} \ne 0$.
\beprop \label{prop4}
 Let  $|\Psi_n^{(1)}\rangle$ and
$|\Psi_n^{(2)}\rangle$ be normalized eigenstates of  $H_1$ and
$H_2$, respectively, such that $A|\Psi_n^{(1)}\rangle \ne 0$,
$A|\Psi_n^{(2)}\rangle \ne 0 $ and the corresponding eigenvalues
are, respectively,  $E_n^{(1)} \ne 0$ and  $E_n^{(2)} \ne 0$. Then
$E_n^{(1)}$ is also an eigenvalue of  $H_2$ associated to the
eigenstate
$$|\Psi_{n-1}^{(2)}\rangle = (E_n^{(1)})^{-1/2}A|\Psi_n^{(1)}\rangle;$$
$E_n^{(2)}$ is also an eigenvalue of  $H_1$ associated to the
eigenstate $$|\Psi_{n+1}^{(2)}\rangle  = (E_n^{(1)})^{-1/2}A^\dag
|\Psi_n^{(1)}\rangle .$$
\enprop
{\bf Proof}: We have  $H_1
|\Psi_n^{(1)}\rangle = E_n^{(1)} |\Psi_n^{(1)}\rangle$. From this, we
deduce that
 $A A^\dag (A|\Psi_n^{(1)}\rangle)= E_n^{(1)} (A|\Psi_n^{(1)}\rangle),$ or \\
\bea \label{g13}
 H_2 (A|\Psi_n^{(1)}\rangle)= E_n^{(1)} (A|\Psi_n^{(1)}\rangle),
 \ena
i.e,   $A|\Psi_n^{(1)}\rangle$ is an eigenstate of  $H_2$ associated to
  the eigenvalue
  $E_n^{(1)}$.
Similarly,  $H_2 |\Psi_n^{(2)}\rangle = E_n^{(2)}
|\Psi_n^{(2)}\rangle$ implies
 $ A^\dag A (A^\dag |\Psi_n^{(2)}\rangle)= E_n^{(2)}
 (A^\dag |\Psi_n^{(2)}\rangle),$
 i.e,
\bea \label{g14}
 H_1 (A^\dag |\Psi_n^{(2)}\rangle)=E_n^{(2)} (A^\dag
 |\Psi_n^{(2)}\rangle).
 \ena
This means that  $A^\dag
 |\Psi_n^{(2)}\rangle$ is an eigenstate of  $H_1$ with the eigenvalue
  $E_n^{(2)}$.
The eigenvalues of  $H_1$ being non degenerate (since we consider
only bound state of  $H_1$), it follows that there exists a unique
normalized eigenstate  $|\Psi_k^{(1)}\rangle$ of $H_1$, up to a
multiplicative constant, corresponding to an eigenvalue $E_k^{(1)}$
such that
 $|\Psi_k^{(1)}\rangle  = c A^\dag |\Psi_n^{(2)}\rangle$.
The normalization constant $c$ is  given by  $c = (E_n^{(2)})^{-1/2}$.
 We have
 \bea \label{g15} |\Psi_k^{(1)}\rangle  = (E_n^{(2)})^{-1/2} A^\dag
|\Psi_n^{(2)}\rangle.
 \ena
\ni  It follows from  (\ref{g15}) that
 \beano
 H_1|\Psi_k^{(1)}\rangle & = &  (E_n^{(2)})^{-1/2}H_1( A^\dag \Psi_n^{(2)})\\
 &  = &
   (E_n^{(2)})^{-1/2} E_n^{(2)} (A^\dag |\Psi_n^{(2)})
 \quad    \   \quad \textrm{from}\quad (\ref{g14})\\
   & = & E_n^{(2)} |\Psi_k^{(1)}\rangle   \quad   \quad   \quad   \quad
    \quad   \quad   \quad    \quad \textrm{from}\quad (\ref{g15}).
 \enano
 Then
$
 H_1|\Psi_k^{(1)}\rangle = E_k^{(1)}|\Psi_k^{(1)}\rangle =
 E_n^{(2)} |\Psi_k^{(1)}\rangle
$. It follows from this that  $E_k^{(1)} = E_n^{(2)}.$ Since
$E_0^{(1)} \ne E_0^{(2)}$ ($E_0^{(1)} =0$ and $E_0^{(2)}\ne 0$),
 a simplest solution of the index equation  is  $k = n+1$.
 Hence
 \bea \label{g16}
\left\{ \ba{rcl}
E_{n+1}^{(1)} & = &E_{n}^{(2)}\\
|\Psi_{n+1}^{(1)}\rangle & = & (E_n^{(2)})^{-1/2} (A^\dag
|\Psi_n^{(2)}\rangle). \ea  \right. \ena One can similarly show from
(\ref{g13}) that
 \bea \label{g17}
\left\{ \ba{rcl}
E_n^{(2)} & = &E_{n+1}^{(1)}\\
|\Psi_{n}^{(2)}\rangle & = & (E_n^{(1)})^{-1/2} (A
|\Psi_{n+1}^{(2)}\rangle). \ea  \right. \ena  \cqfd
\ni It follows from these propositions that the eigenvalues of   $H_1$
and $H_2$ are positive definite
 ($E_n^{1,2} \ge 0$),  and the partner  Hamiltonians are   isospectral, i.e.,
 they have almost the same  energy eigenvalues,  except for the ground state
 energy of $ H_1$
 which is missing in the spectrum  of  $H_2$.
The spectra are linked as
\cite{Cooper}:
 \bea  \label{g18}
 E_n^{(2)} &= & E_{n+1}^{(1)},\quad
E_0^{(1)} = 0, \quad  n= 0,1,2,\ldots ,\cr { \Psi}_n^{(2)} &=
&\left[ E_{n+1}^{(1)}\right]^{(-1/2)}A { \Psi}_{n+1}^{(1)},\cr
        { \Psi}_{n+1}^{(1)} &= &\left[ E_{n}^{(2)}\right]^{(-1/2)}A^{\dag}
{ \Psi}_n^{(2)}.
 \ena
 Hence, if the eigenvalues and eigenfunctions of  one of the partner, say $H_1$,
 are known, one can immediately derive the eigenvalues and
eigenfunctions of  ${H}_2$.\\\\
However, the above relations (\ref{g18}) only give the relationship between the
 eigenvalues and eigenfunctions of  the two partner Hamiltonians,   but do not allow to determine their spectra.
A condition of an exact solvability is known as the shape invariance condition;
  that is,  the pair of SUSY partner potentials ${V}_{1,2}$ are similar
  in shape and differ only in the  parameters that appear in them. 
Gendenshtein states the shape invariance condition as\cite{Gend, Cooper}
 \bea \label{g19}
  {V}_2(x; a_1) = {V}_1(x;a_2) + {\cal R}(a_1),
  \ena
  where $a_1$ is a set of parameters  and $a_2$ is a
   function of $a_1,$  $(a_2 = f(a_1)),$
  and ${\cal R}(a_1)$ is the non-vanishing remainder independent of $x$. In such a case,
  the eigenvalues  and the eigenfunctions of ${H}_1$ can explicitly  be deduced \cite{Gend}. 
If this  Hamiltonian   $H_1$ has $p \,(p\ge 1)$ bound states with
 eigenvalues
  $E_n^{(1)}$, and eigenfunctions  $\Psi_n^{(1)}$ with $0 \le n \le p-1$,
 the starting point of constructing the spectra is to  generate a hierarchy of  $(p-1)$ Hamiltonians
 $H_2, \ldots H_p$  such that the $m$'th  member of the hierarchy
 $(H_m)$ has the same spectrum as  $H_1$ except that the first $m-1$
 eigenvalues of $H_1$ are missing in the spectrum of  $H_m$\cite{Cooper}.
 In order $m$, $(m = 2, 3, \ldots p)$, 
we have partner  Hamiltonians
  \beano
 H_m(x; a_1)& =&  A_m^\dag (x; a_1) A_m(x; a_1)  + E_0^{(m)} = -{d^2\over dx^2} + V_m(x; a_1),\\
  H_{m+1}(x; a_1)& = & A_m(x; a_1) A_m^\dag(x; a_1)  + E_0^{(m)} = -{d^2\over dx^2} + V_{m+1}(x; a_1),
\enano the spectra of which  are related as
 \beano E_n^{(m+1)} = E_{n+1}^{(m)},
\qquad \Psi_n^{(m+1)} = (E_{n+1}^{(m)} - E_0^{(m)})A_m \Psi_{n+1}^{(m)}. 
\enano
In
terms of the spectrum of  $H_1$ we have
 \bea \label{g20}
E_n^{(m)} & = & E_{n+1}^{(m-1)} = E_{n+2}^{(s-2)} = \cdots = E_{n+m-1}^{(1)}\\
\Psi_n^{(m)} & = & (E_{n+m-1}^{(1)} - E_{m-2}^{(1)})^{-1/2}\cdots
 (E_{n+m-1}^{(1)} - E_0^{(1)})^{-1/2}A_{m+1}\cdots A_1 \Psi_{n+m-1}^{(1)}(x; a_1).\nonumber
\ena
\betheo
The eigenvalues of $H_1$ are given by \cite{Gend, Cooper}
 \bea \label{g21}
  E_n^{(1)} =\sum_{k =1}^n {\cal R}(a_k). 
  \ena
\entheo
{\bf Proof}:
\noindent Consider the partner  Hamiltonians $H_m$ and $H_{m+1}$
of the hierarchy of Hamiltonians constructed from  $H_1$. If the
partner potentials are shape invariant, we can write
   \beano
   V_{m+1}(x; a_1) & = & V_m(x; a_2) + {\cal R}(a_1) \\
   & =  & V_{m-1}(x; a_3) + {\cal R}(a_2) +{\cal R}(a_1)\\
   & = & V_{m-2}(x; a_4) + {\cal R}(a_3) +{\cal R}(a_2) +{\cal R}(a_1)\\
   &  & \vdots \\
   &  = & V_2(x; a_m) + {\cal R}(a_{m-1})+ {\cal R}(a_{m-2}) + \cdots +
   {\cal R}(a_1)\\
   & = & V_1(x; a_{m+1}) + \sum_{k=1}^m {\cal R}(a_k).
   \enano
It follows from the above that
   $H_m(x; a_1) = H_1(x; a_m) +   \dis\sum_{k=1}^{m-1} {\cal R}(a_k)$.
Hence
   $E_0^{(m)} =  \dis \sum_{k=1}^{m-1} {\cal R}(a_k)$.
From equation  (\ref{g18}), $E_0^{(m)} = E_{m-1}^{(1)}$. Then
   $E_{m-1}^{(1)} =  \dis \sum_{k=1}^{m-1} {\cal R}(a_k)$,
i.e, 
$ E_n^{(1)} =   \dis \sum_{k=1}^n {\cal R}(a_k)$ .
 \cqfd
\betheo
The normalized eigenfunctions of $H_1$  are given by \cite{Dab}
 \bea \label{g22}
{\Psi}_n(x; a_1) = \left\{\prod_{k=1}^n\left(\sum_{p=1}^k {\cal R}(a_p) \right)
  \right\}^{-1/2}  A^\dag(x; a_1) \cdots A^\dag(x; a_{n})\Psi_0^{(1)}(x;a_{n+1}).
 \ena
\entheo
{\bf Proof}:
\noindent From the shape invariance condition  (\ref{g19}), we deduce the
following relation between the eigenfunctions of the partner
Hamiltonians $H_1$ and $H_2$
 \bea \label{g22}
 \Psi_n^{(2)}(x; a_1) = \Psi_n^{(1)}(x; a_2)
 \ena
We know from (\ref{g17}) that
 \beano
 \Psi_{n+1}^{(1)}(x; a_1) & = & (E_n^{(2)})^{-1/2}A^\dag(x; a_1) \Psi_n^{(2)}(x; a_1)\\
 & = &   (E_n^{(2)})^{-1/2}A^\dag(x; a_1) \Psi_n^{(1)}(x;a_2) \quad \textrm{from}\quad (\ref{g22})\\
 & = &   (E_n^{(2)})^{-1/2} (E_{n-1}^{(2)})^{-1/2}
          A^\dag(x; a_1)A^\dag(x;a_2) \Psi_{n-1}^{(2)}(x;a_3)\\
 & = & \vdots \\
 & = & (E_n^2)^{-1/2} \cdots (E_0^2)^{-1/2}
 A^\dag(x; a_1)\cdots A^\dag(x; a_{n+1}) \Psi_0^{(1)}(x;a_{n+2}).
 \enano
It deduces  from  above equations that
 \bea \label{function}
 \Psi_n^{(1)}(x; a_1)  = \left\{\prod_{k=1}^n\left(\sum_{p=1}^k {\cal R}(a_p) \right)
  \right\}^{-1/2}  A^\dag(x; a_1) \cdots A^\dag(x; a_{n})\Psi_0^{(1)}(x;a_{n+1}).
 \ena
\cqfd \ni
The shape invariance condition (\ref{g19}) can be rewritten in terms
of the factorization operators
defined in equations  (\ref{g4})-(\ref{g7}),
\bea \label{g23}
 {A}(a_1){A}^\dag(a_1) = {A}^\dag(a_2){A}(a_2) +  {{\cal R}(a_1)} ,
\ena
where  $a_2$ is a function of $a_1$. Here,  we  consider only the translation class of
 shape invariance
 potentials, that is the case where the parameters $a_1$ and $a_2$ are related as
   $a_2 = a_1 + \eta$  \cite{Dab} and the potentials are known in closed form. The
   scaling class \cite{Suk} is not treated here since the potentials, in this case, can only
    be written as Taylor expansion.\\\\
Introducing  a reparametrization  operator $T_\eta$ defined as 
\bea\label{g24}
T_\eta : {\cal H} \longrightarrow {\cal H} \qquad T_\eta \Phi(x;a_1) := \phi(x; a_1 + \eta) = \Phi(x; a_2)
\ena
that replaces $a_1$ with $a_2$ in a given
operator \cite{Bal}
\bea \label{g25}
{T}_\eta {{\cal O}(a_1)} T^{-1}_\eta =  {{\cal O}(a_1 + \eta )} :=  {{\cal O}(a_2)} ,
\ena
and the  operators
\bea \label{g26}
{B}_-, {B}_+ : {\cal H} \longrightarrow {\cal H} \quad {B}_+  = {A}^\dag(a_1) {T}_\eta ,\quad
{B}_-  = {T}^\dag_\eta {A}(a_1) ,
\ena
with the domains
\bea \label{g27}
{\cal D}({B}_-) & = & \left\{ u \in {\cal H}, \quad  v =  u'+  W u
 \in {\cal H} \quad  \textrm{and} \quad  {T}^\dag_\eta v  \in {\cal H} \right\} \\
{\cal D}({B}_+) & = & \bigg\{ u \in {\cal H}, \quad  v =  {T}_\eta u \in {\cal H}\quad  \textrm{and} \quad   -v'+  W v
 \in {\cal H}   \bigg\}.
\ena
The Hamiltonian factorizes in terms of the new operators as follow:
\bea \label{g28}
H - E_0 = H_1 = A^\dag(a_1) A(a_1) = B+ B_-
\ena
where
\bea \label{g29}
[B_-, B_+] & = & {\cal R}(a_0)\quad, \quad 
B_-\lv \Psi_0\rs  = 0 \pt
\ena
The states $B_+^n\lv\Psi_0\rs$ are eigenfunctions of $ H$ with eigenvalues $E_n$, ie, 
\bea \label{g30}
H(B_+^n \lv \Psi_0\rs) & = & \underbrace{\left[\sum_{k = 1}^n {\cal R}(a_k)\right]}_{E_n} B_+^n \lv \Psi_0\rs
\ena
$B_{\pm}$  act as raising and lowering operators: 
\bea \label{g31}
B_+\lv \Psi_n\rs  & = & \sqrt{E_{n+1}} \lv\Psi_{n+1}\rs \quad, \quad 
B_-\lv \Psi_n\rs = \sqrt{ {\cal R}(a_0) + E_{n-1}} \lv \Psi_{n-1}\rs \pt
\ena
To define shape-invariant potential coherent states, 
 Balantekin {\it et al} \cite{Alex} introduced the right inverse of $B_-$ as: $B_- B_-^{-1}  = \id$
and the left inverse $H^{-1}$ of $H$ such that: $H^{-1} B_+ = B_-^{-1}$. 
The SIPCS defined by 
\bea \label{g32}
\lv z\rs = \sum_{n = 0}^n (z B_-^{-1})^n \lv\Psi_0\rs
\ena
are   eigenstates of the lowering operator $B_-$: 
\bea \label{g33}
B_-\lv z\rs & = & z \lv z\rs\pt
\ena
A generalization of the SIPCS (\ref{g32}) was done as \cite{Alex}: 
\bea \label{g38}
\lv z ; a_j\rs = \sum_{n=0}^\infty \left\{ z Z_j B_-^{-1}\right\}^n \lv \Psi_0\rs, \quad z, Z_j \in \IC
\ena
where  $Z_j \equiv Z_{(a_j)} \equiv Z(a_1, a_2, \ldots) $.\ 
Observing that  $B_-^{-1} Z_j = Z_{j+1}\ B_-^{-1} $ 
and from 
\bea \label{g39}
Z_{j-1}  = T^\dag(a_1) Z_j T(a_1)
\ena
one can readily show that 
\bea \label{g40}
(z Z_j B_-^{-1})^n = z^n \prod_{k = 0}^{n - 1} Z_{j+k} B_{-}^{-n}\pt
\ena
Using (\ref{g40}), one can  straightforwardly deduce  that (\ref{g38}) are  eigenstates of $B_-$: 
\bea \label{g41}
B_- \lv z; a_j\rs = z Z_{j -1} \lv z;a_j\rs \pt
\ena
Observing that 
\bea \label{g42}
B_-^{-n}\lv \Psi_0\rs = C_n\lv\psi_n\rs , \quad C_n = \left[\prod_{k = 1}^n\left(\sum_{s = k}^n {\cal R}(a_s) \right) \right]^{-1/2}
\ena
and using  (\ref{g42}), 
the
 normalized form of the CS (\ref{g38}) can be obtained as: 
\bea \label{g43}
\lv z; a_r\rs = {\cal N}(|z|^2; a_r)\sum_{n = 0}^\infty {z^n \over h_n(a_r)}\lv \Psi_n\rs \virg
\ena
where we used the shorthand notation $a_r \equiv [{\cal R}(a_1), {\cal R}(a_2), \ldots,{\cal R}(a_n)\,;\,a_j, a_{j+1},\ldots, a_{j+n -1} ]$. 
The expansion coefficient $h_n(a_r)$ and the normalization constant $ {\cal N}(|z|^2; a_r)$ are:
\bea \label{g44}
 h_n(a_r) = \dis\frac{\sqrt{\dis \prod_{k = 1}^n \left(\sum_{s = k}^n {\cal R}(a_s) \right)}}
{\dis \prod_{k = 0}^{n - 1} Z_{j+k}} \quad 
\textrm{for} \  n \ge 1,\   h_0(a_r) = 1, \quad{\cal N}(x; a_r) = \left[\sum_{n = 0}^\infty \, {x^n \over |h_n(a_r)|^2 } \right]^{-1/2}\pt
\ena
It is shown \cite{Alex} that these states (\ref{g38}) fulfill the standard properties of label continuity, overcompleteness, temporal stability and action identity. 
\section{Construction of photon-added coherent states for shape invariant systems }
In this section, a construction of PA-SIPCS \cite{SAH_prep}, and  their physical and  mathematical properties are presented.
\subsection{Definition of the PA-SIPCS}
\ni Let  ${\mathfrak H}_m $  be the  Hilbert subspace  of $\mathfrak H$ defined  as follows:
\bea \label{c1}
 {\mathfrak H}_m := span\left\{\lv\Psi_{n+m}\rs\right\}_{n,m \ge 0}.
\ena
By successive application of the raising operator $B_+$ on  the generalized SIPCS 
 (\ref{g33}), we can obtain 
photon-added shape-invariant potential CS (PA-SIPCS)  denoted by  $\lv z ; a_r \rs_m$:
\bea \label{c2}
\lv z ; a_r \rs_m & :=  & (B_+^m)\lv z ; a_r \rs
\ena
 where $m$ is a positive integer standing for  the number of added quanta or  photons. \\
It is worth mentioning that  the first $m$ 
eigenstates $\lv\Psi_n\rs $, $n =0,1, \dots, m-1$ are absent from the wavefunction $\lv z ; a_r \rs_m \in {\mathfrak H}_m.$ 
Therefore, from the 
orthonormality relation  satisfied by the states $\lv\Psi_n\rs, $   the  overcompleteness relation fulfilled by 
the  identity operator on  ${\mathfrak H}_m,$ denoted  by 
$\id_{{\mathfrak H}_m}$, is to  be written as \cite{Penson, Popov}
\bea \label{c3}
 \id_{{\mathfrak H}_m} = \sum_{n = m}^\infty \lv\Psi_{n} \rs \li \Psi_{n}\rv = \sum_{n = 0}^\infty \lv\Psi_{n+m} \rs \li \Psi_{n + m}\rv.
\ena 
Here, $\id_{{\mathfrak H}_{m}}$ is only required to be a bounded positive operator with a densely defined inverse 
\cite{Ali95}.\\\\
\ni From (\ref{g25}) and  using the relations $B_+ {\cal R}(a_{n-1}) = {\cal R}(a_n) B_+$
and $B_+\lv \Psi_n\rs = \sqrt{E_{n+1}} \lv \Psi_{n+1}\rs$,
we obtain 
the  PA-SIPCS  as:
\bea \label{c4}
\lv z;a_r\rs_m = {\cal N}_m(|z|^2; a_r) \sum_{n = 0}^\infty {z^n \over K_n^m(a_r) } \lv \Psi_{n+m}\rs
\ena
where the the expansion coefficient takes the form:
\bea \label{c5}
K_n^m(a_r)  & = & \dis \frac{ \dis \left[\prod_{k = m+1}^{n+m} \left(\sum_{s = k}^{n+m} {\cal R}(a_s) \right) \right]^{1/2} }
{\dis  \left[\prod_{k = m}^{n+m-1} Z_{j+k}\right].
\left[ \prod_{k = 1}^m \left(\sum_{s=k}^{n+m}{\cal R}(a_s)\right) \right]^{1/2}} 
 \virg
\ena
and the  normalization constant ${\cal N}_m(|z|^2; a_r)$ is given by:
\bea \label{c6}
{\cal N}_m(|z|^2; a_r) = \left( \sum_{n = 0}^\infty {|z|^{2n} \over |K_n^m(a_r)|^2 }\right)^{-1/2} \pt
\ena
 The inner product of two different PA-SIPCS $\lv z; a_r\rs_m$ and $\lv z'; a_r\rs_{m'}$ 
\bea \label{c7}
 {}_{m'}\li z';a_r\right.\lv z; a_r\rs_m & = &  {\cal N}_{m'}(|z'|^2; a_r) {\cal N}_m(|z|^2; a_r) 
\sum_{n,n' = 0}^\infty { {{z'}^\star}^{n'} z^n  \over {K_{n'}^{m'}}^\star(a_r) K_n^m(a_r)} 
\li \Psi_{n'+m'} \right. \lv  \Psi_{n+m} \rs
\ena 
does not vanish. Indeed, due to the orthonormality of the eigenstates $|\Psi_n\rangle$, the inner product (\ref{c7}) can be rewritten as
\bea \label{c8}
 {}_{_{m'}}\li z';a_r\right.\lv z; a_r\rs_m & = &  {\cal N}_{m'}(|z'|^2; a_r) {\cal N}_m(|z|^2; a_r) {z'^\star}^{(m - m')}
\sum_{n = 0}^\infty {({z'}^\star z)^n \over  {K_{n +m-m'}^{m'}}^\star(a_r)\ K_n^m(a_r)},
\ena
 showing that the PA-SIPCS are not mutually orthogonal. 
\subsection{Label continuity} 
\ni  In the Hilbert  space ${\mathfrak H}$, the PA-SIPCS $|z,a_r\rangle_m$ are labeled by $m$ and $z$. The label continuity condition can then be stated as:
{\footnotesize
\bea \label{c9}
 |z -z'| \to 0 \ \textrm{and} \  |m - m'|\to 0 \Longrightarrow  
|||z,a_r\rangle_m - |z',a_r\rangle_{m'}||^2 = 2\left[1 - {\cal R}e \left({}_{_{m'}}\li z';a_r\right.\lv z; a_r\rs_m\right)\right]  \to 0.
\ena}
This is satisfied by the states  $|z,a_r\rangle_m$, since from Eqs. (\ref{c6}, \ref{c8}), we see that
\bea \label{c10}
m \to m' \quad  \textrm{and} \quad  z \to z' \Longrightarrow {}_{_{m'}}\li z';a_r\right.\lv z; a_r\rs_m \to 1.
\ena
Therefore the PA-SIPCS  $|z,a_r\rangle_m$ are continuous in their labels.
\subsection{Overcompleteness}
\ni  We check the realization of the resolution of identity in the Hilbert space  (\ref{c1}) with the identity operator defined as (\ref{c3}):
\bea \label{c11}
 \int_{{\IC}}\  d^2z\, \lv z; a_r\rs_m\, \omega_m(|z|^2;a_r)\, {}_{m}\li z;a_r\rv  = \id_{{\mathfrak H}_m}. 
\ena
Inserting the definition (\ref{c4}) of the PA-SIPCS $\lv z; a_r\rs_m$ into Eq. (\ref{c11}) yields, 
after taking  the angular integration of the diagonal matrix elements: 
\bea \label{c12}
  \int_0^\infty dx\  x^n\  {\cal W}_m(x ; a_r) =  |K_n^m(a_r)|^2, \quad  
\quad \textrm{with} \quad 
 {\cal W}_m(x ; a_r) = \pi {\cal N}_m^2(x; a_r)\  \omega_m(x ; a_r).
\ena
Therefore, the weight function $\omega_m$ is related to the undetermined  moment distribution ${\cal W}_m(x ; a_r)$, which is the  solution of the Stieltjes moment problem with the moments given by $ |K_n^m(a_r)|^2$.
 In order to use Mellin transformation, we can  rewrite (\ref{c12}) as 
\bea \label{c13}
  \int_0^\infty  dx\  x^{n+m}\ g_m(x ; a_r) =  |K_n^m(a_r)|^2, \quad \textrm{where} \quad 
g_m(x; a_r) = \pi {{\cal N}_m^2(x; a_r)} x^{-m} \ \omega_m(x; a_r)\pt
\ena
By performing the variable change  $n + m \to s -1,$ \  Eq. (\ref{c13}) becomes:
\bea \label{c14}
\int_0^\infty dx x^{s-1} g_m(x;a_r) =  |K_s^m(a_r)|^2 \pt
\ena
Comparing this relation with the Meijer's G-function and the Mellin inversion theorem \cite{Mathai} 
{\small
\bea \label{c15}
  \int_0^\infty dx \, x^{s-1} G_{p,q}^{m,n} 
\left(\alpha x\lv \ba{c} 
a_1,... , a_n  ; a_{n+1},  ..., a_p \\
b_1,... , b_m  ; b_{m+1},  ..., b_q 
\ea \right.\right)
={1 \over \alpha^s}\ {\dis \prod_{j = 1}^m \Gamma(b_j+s) \prod_{j = 1}^n \Gamma(1 - a_j - s) \over  
\dis \prod_{j = m +1}^q \Gamma(1 - b_j -s) \prod_{j = n + 1}^p \Gamma(a_j + s)} \virg
\ena
}
we see that  if  $ |K_s^m(a_r)|^2$  in the above relation can be expressed 
in terms of Gamma functions, then  
$ g_m(x;a_r)$ can be identified as the Meijer's G- function.
\subsection{Thermal statistics} 
\ni In quantum mechanics, the density matrix, generally denoted by $\rho$,  is an important  tool for characterizing   the probability distribution on the states 
of a physical system. For example, it is useful 
for examining the physical and chemical properties of a system (see
 \cite{Popov}, \cite{Aremua2} and references listed therein). Consider a quantum gas of the system in the thermodynamic
equilibrium with a reservoir at temperature $T$, which satisfies a quantum canonical
distribution. The corresponding normalized density operator is given, in the Hilbert space $ {\mathfrak H}_m := span\left\{\lv\Psi_{n+m}\rs\right\}_{n,m \ge 0}$,  as 
\bea\label{thermal00}
 \rho^{(m)} = \frac{1}{Z}\sum_{n=0}^{\infty} e^{-\beta E_n}|\Psi_{n + m} \rangle 
  \langle  \Psi_{n + m}|,
\ena
where in the exponential $E_n$ is the eigen-energy, and  the partition function $Z$ is taken as the normalization constant.\\
The diagonal elements of  $\rho^{(m)}$, essential for our purpose, 
also known as the $Q$-distribution or Husimi's distribution,  are derived  in the  PA-SIPCS basis as 
\bea\label{thermal01}
{}_{{m}}\li z;a_r| \rho^{(m)} | z; a_r\rs_m   = \frac{{\cal N}_m^2(|z|^2 ; m) }{Z}
\sum_{n = 0}^\infty \frac{|z|^{2n} }{ |K_n^p(a_r)|^2 } e^{-\beta E_n}.
\ena
The normalization of the density operator leads to
\bea\label{thermal02}
\mbox{Tr} \rho^{(m)}  =  \int_{\IC}  d^2 z \, \omega_m(|z|^2; a_r)  \, _{m}\langle z;a_r| \rho^{(m)} |z;a_r\rangle_m =  1.
\ena
The diagonal expansion of the normalized canonical density operator  over the  PA-SIPCS projector   is
\bea\label{thermal03}
 \rho^{(m)}  = \int_{\IC} d^2 z \, \omega_m(|z|^2; a_r) |z;a_r\rangle_m P(|z|^2) \, _{m}\langle z;a_r|,
\ena
where the $P$-distribution  function  $P(|z|^2)|$ satisfying the normalization to unity condition 
\bea\label{thermal04}
 \int_{\IC} d^2 z \, \omega_m(|z|^2; a_r) P(|z|^2) = 1 
\ena
must  be determined.\\
Thus, given an observable $\mathcal O$, one obtains the expectation value,  i. e., the thermal average given by
\bea{\label{obaverage00}}
\langle \mathcal O \rangle_{m} = Tr ( \rho^{(m)}   \mathcal O) =  \int_{\IC} d^2 z \, \omega_m(|z|^2; a_r)
P(|z|^2) \,_{m}\langle z;a_r|\mathcal O |z;a_r\rangle_{m}.
\ena
One can check that for a PA-SIPCS (\ref{c4}) the expectation values of the operator $N :=  B_+ B_-$\cite{SAH_prep}  are:
\bea \label{c20}
 \li N \rs = {\cal N}_m^2(|z|^2 ; a_r) \sum_{n = 0}^\infty E_{n + m}\ {|z|^{2n}\over  |K_n^m(a_r)|^2 } \ , \ 
\li N^2 \rs =  {\cal N}_m^2(|z|^2 ; a_r) \sum_{n = 0}^\infty E_{n + m}^2\ { |z|^{2n} \over |K_n^m(a_r)|^2 }\pt
\ena 
Using (\ref{c20}), the  pseudo-thermal expectation values of the  operator $N$ and of its square $N^2, $
given by $\langle   N \rangle^{(m)} = Tr (\rho^{(m)}N)$ and 
$\langle N^2 \rangle^{(m)} = Tr (\rho^{(m)} N^2)$, respectively, 
allow to obtain 
the  thermal intensity correlation function as follows:
\bea\label{thermcor}
(g^2)^{(m)}  =  \frac{\langle N^2 \rangle^{(m)} - \langle   N \rangle^{(m)}}{\left(\langle   N \rangle^{(m)}\right)^{2}}.  
\ena
Then, the   thermal analogue of the Mandel parameter given by
\bea\label{thermcor00}
 Q^{(m)}  =  \langle  N \rangle^{(m)}\left[(g^2)^{(m)}-1\right] 
\ena
is deduced.
\section{P\"oschl-Teller potential}
\ni 
 Consider the family of potentials 
\bea \label{p1}
V_{l, l'}(x) = \left\{\ba{l}
\dis {1\over 4 a^2}\left[{l(l - 1) \over \sin^2 u(x)} - {l'(l' - 1) \over \sin^2 u(x)}  \right] - 
\dis {(l + l')^2 \over 4 a^2}, \ u(x) = {x\over 2 a}, \  \quad 0 < x < \pi a  \\\\
\infty, \quad x \le 0, x \ge \pi a
\ea \right.
\ena 
 of continuously indexed parameters $l, l'$. This class of  potentials  called  P\"oschl-Teller  potentials of first type (PT-I), intensively studied in  \cite{Antoine, Bergeron, Sama, Kinani},  is closely related to other classes of potentials, widely used in molecular physics, namely  
(i) the symmetric P\"oschl-Teller potentials well $(l = l' \ge 1)$, 
(ii) the Scarf potentials $\half < l' \le 1$ \cite{Scarf}, 
(iii) the modified P\"oschl-Teller potentials which can be obtained by replacing the trigonometric functions by their hyperbolic counterparts \cite{Poschl-Teller, Daskalo},
(iv) the Rosen-Morse potential which is the symmetric modified P\"oschl-Teller potentials \cite{Rosen}.\\\\
Let us define the corresponding Hamiltonian operator $H_{l,l'}$ with the action 
\bea \label{p2}
H_{l,l'}\phi := \left(-{d^2 \over d x^2} + 
{1\over 4 a^2}\left[{l(l - 1) \over \sin^2 (x/2a)} - {l'(l' - 1) \over \sin^2 (x/2a)}  \right] - 
\dis {(l + l')^2 \over 4 a^2}
\right)\phi \quad \textrm{for}\quad \phi \in {\cal D}_{H_{l,l'}}
\ena
in the suitable Hilbert space ${\cal H} = L^2((0,\pi a), dx)$. 
 ${\cal D}_{H_{l,l'}}$ is  the domain of definition of $H_{l,l'}$.
We consider here the case where $l, l' \ge 3/2$, then the operator ${H}_{l,l'}$ is in the limit point case at both ends $x = 0$, $\pi a,$
therefore it is essentially self-adjoint.  In this case (see  \cite{Antoine, Bergeron} for more  details) the P\"oschl-Teller Hamiltonian can be defined as the self-adjoint operator $H_{l,l'}$ in $L^2([0, \pi a], dx)$ acting as in (\ref{p2}), on the dense domain 
\bea \label{p3}
{\cal D}_{H_{l, l'}} = \bigg\{\phi \in AC^2(0, \pi a)| \ 
\left( {1\over 4 a^2}\left[{l(l - 1) \over \sin^2 (x/2a)} - {l'(l' - 1) \over \sin^2 (x/2a)}  \right] - 
\dis {(l + l')^2 \over 4 a^2}\right) \phi \in L^2([0, \pi a], dx), 
 \cr
\quad
 \phi(0) = \phi(\pi a) = 0 \bigg\}.
\ena 
with $AC^2(0,\pi a) = \left\{\phi \in ac^2(0, \pi a) : \phi' \in {\cal H} \right\},$ 
where $ac^2(0,\pi a)$ denotes the set of absolutely continuous functions with abolutely continuous derivatives.\\\\
PT-I potentials are SUSY and fullfill the property of shape invariance \cite{Cooper}. 
 Their superpotentials   are:
\bea \label{p4}
 W(x; l, l') & = & -{1\over 2a}\left[l \cot(u(x)) - l' \tan (u(x)) \right].
\ena
One can define the first differential operators $A, A^\dag$ that factorize the Hamiltonian operator in
 (\ref{p2})  as: 
\bea \label{p5}
A := {d\over dx} + W(; l, l'), \quad A^\dag := -{d\over dx} + W(x; l, l')
\ena
with  the domains: 
\bea \label{p6}
{\cal D}(A) & = &  \{\psi \in ac[0, \pi a], \quad (\psi' + W(x; l, l') \psi) \in {\cal H}\}\\
{\cal D}(A^\dag) & = & \left\{\phi  \in ac[0, \pi a]| \exists
 {\tilde \phi} \in {\cal H}: [\psi(x) \phi(x)]_0^a = 0,
\li A\psi, \phi\rs =   \li\psi, {\tilde \phi}\rs, \forall \psi \in {\cal D}(A)\right\}
\ena
with $A^\dag \phi = {\tilde \phi}$.
The partner potentials $V_{1,2}$ satisfy the following shape invariance relation: 
\bea\label{p10}
V_2(x, l, l') = V_1(x, l + 1, l' + 1) + {1\over a^2} (l + l' + 1).
\ena
The potential parameters $a_1 \equiv (l, l')$ and $a_2 \equiv (l+1, l'+1)$   are related as 
\bea \label{p11}
a_2 = a_1 + 2,
\ena
while the remainder  in the shape invariant condition
 (\ref{g14}) is  ${\cal R}(a_1)  =  {1 \over a^2} (l + l' +1) $. 
Then the  products in terms of the quantity ${\cal R}(a_s)$ in the numerator and denominator of the coefficient $K_n^m(a_r)$, see Eq. (\ref{c5}), can be read, respectively, as:
\bea \label{p12}
\prod_{k = m+1}^{n + m} \left(\sum_{s=k}^{n+m} {\cal R}(a_s) \right) & = &  \lambda^{2n}\frac{\Gamma(n+1) \Gamma(2 n+ 2 m + 2 \rho)}
                                                      {\Gamma(n + 2m + 2\rho)} \\\label{p13}
\prod_{k =1}^{m}\left(\sum_{s=k}^{n+m} {\cal R}(a_s) \right)& = &  \lambda^{2m}\frac{\Gamma(n+m+1) \Gamma(n+ 2 m + 2 \rho)}
                                                      {\Gamma(n+1)\Gamma(n+ m + 2\rho)} 
\ena
where we set $\lambda = {1\over a}$ and $2 \rho = l + l', \rho \ge 3/2$. 
The explicit form of the expansion coefficient $K_n^m(a_r)$ depends on the choice of the functional 
${\cal Z}_j$. 
\subsection{First  choice of the functional ${\cal Z}_j$} 
\ni First we define  the  functional ${\cal Z}_j$ as 
 ${\cal Z}_j  = e^{-i \alpha {\cal R}(a_1)}$, then 
 we obtain 
\bea \label{p15}
\prod_{k = m}^{n+m-1} {\cal Z}_{j+k} = e^{_ i \alpha E_n},
 \quad  E_n =  \lambda^2 n(n + 2 \rho).
\ena 
 Inserting this relation  and the  results (\ref{p12}) and (\ref{p13}) in  (\ref{c5}),
 we obtain the expansion coefficient as:
\bea \label{p16}
K_n^m(a_r) & = &  \lambda^{n-m}\ \sqrt{\frac{\Gamma(n+1)^2\ \Gamma(2n+2m+2 \rho)\ \Gamma(n+ m +2 \rho)}
{\Gamma(n+m+1) \Gamma(n+ 2 m + 2 \rho)^2}}\, e^{i\alpha E_n}.                                                   
\ena
\ni {\it (i) Normalization} \\\\
The normalization factor, in terms of the generalized hypergeometric functions $_3F_4$, can  readily be 
 deduced from  (\ref{c6}) as:
{\small
\bea \label{p17}
{\cal N}_m(|z|^2; a_r) &=  & \left[ 
\dis {\lambda^{2m}\Gamma(m+1)\Gamma(2m + 2 \rho)\over \Gamma(m + 2 \rho)}\,
 {}_3F_4\left(\ba{ccl} m+1, 2m+2\rho,2m+2\rho & ;& \\
  1, m+\rho,m+2\rho, m+\rho + 1/2& ;&  {|a z|^2\over 4}\ea\right) \right]^{-1/2} \pt
\ena}
 In terms of Meijer's G-function, we have:
{\footnotesize
\bea \label{p18}
{\cal N}_m(|z|^2; a_r) & = &
\left[  
\dis {\lambda^{2m} \Gamma(m +  \rho) \Gamma(m +  \rho + \half) \over \Gamma(2m + 2 \rho)}\, 
 G_{3,5}^{1,3}\left(-{|a z|^2\over 4}\left|\ba{l}
                                    -m, 1 - 2 m - 2\rho , 1 - 2 m - 2\rho   \\ 
                      0 , 0, 1-m - \rho, 1-m - 2\rho, 1/2 -m - \rho \ea \right.\right) \right]^{-1/2}.
\ena}
 The explicit form of these PA-SIPCS are:
\bea \label{p19}
  \lv z; a_r\rs_m = {\cal N}_m(|z|^2; a_r) \lambda^m 
\sum_{n = 0}^{\infty} \sqrt{\frac{\Gamma(n+m+1) \Gamma(n+ 2 m + 2 \rho)^2}
                   {\Gamma(n+ m + 2 \rho)\Gamma(2n+2m+2 \rho)}}\,  {(az)^{n}\over n!} \lv n+m\rs
\ena
defined on the whole complex  plane. 
  For $m = 0$, we recover the expansion coefficient  and the normalization factor  obtained in \cite{Alex} for the generalized SIPCS:
{
\bea \label{p20}
 K_n^0 & = & \lambda^n\sqrt{\Gamma(n+1) \Gamma(2\rho + 2n) \over \Gamma(2\rho + n)}\, e^{i \alpha E_n} = h_n(a_r), \cr
 {\cal N}_0(|z|^2; a_r) & = & 
\left[{}_1F_2\left(2\rho; \rho, \rho + 1/2 ; {|az|^2\over 4}\right)\right]^{-1/2} = {\cal N}(|z|^2; a_r)
\pt
\ena}
\begin{figure}[htbp]
\begin{center}
\includegraphics[width=7.5cm]{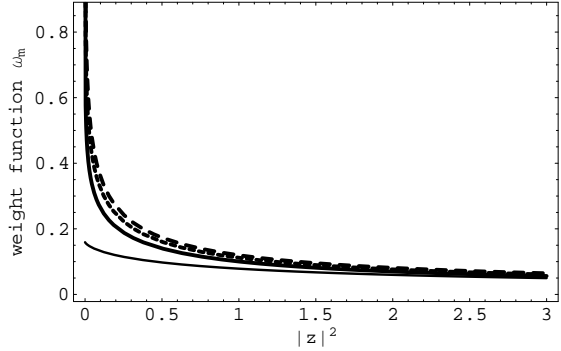}
\end{center}
\ni \caption[]{ 
Plots of the  weight function (\ref{p22}) of  the
 PA-SIPCS (\ref{p19}) versus  $|z|^2$ with the potential parameters $\rho =2, \lambda = 1$,  for different  values of 
 the photon added number $m$  with 
 $m = 0$ (thin solid  line), $m = $ (solid line), $m = 3$ (dot line), and $m = 4$ (dashed line). }
\end{figure}
\newline
\ni {\it (ii) Non-orthogonality} \\\\
The inner product of two different PA-SIPCS $\lv z; a_r\rs_m$ and $\lv z';a_r\rs_{m'}$ follows from Eq (\ref{c8}):
{\small
 \beano
_{m'}\li z';a_r\rv \left. z; a_r\rs_m =\chi(z',z, m, m',\rho)
_3F_4\left(\ba{ccl} m +1,2m+2\rho,m+m'+2\rho &;&\\
 m - m'+ 1, m+2\rho, m+\rho, m+\rho+\half &; & 
{a^2{ z'}^\star z\over 4}\ea \right)
\enano
}
where 
 $ \chi(z',z, m, m',\rho) = {\cal N}_{m'}(|z'|^2;a_r)\ {\cal N}_m(|z|^2;a_r) \ {{ z'}^\star}^{(m-m')} \lambda^{(m+m')}
 {\Gamma(m +1)  \Gamma(m +m'+2\rho) \over \Gamma(m -m'+1)  \Gamma(m +2\rho)}.$ \\\\
\ni {\it (iii) Overcompleteness} \\\\
The non-negative weight function $\omega_m(|z|^2;a_r)$ is related to the  function $g_m$  satisfying  (\ref{c13}):
\bea \label{p21}
 \int_0^\infty dx\,x^{n+m}\, g_m(x;a_r) = \xi(x,n,m,\rho)\,
{\Gamma(n+1)^2 \Gamma(n + m + 2\rho) \Gamma(n + m + \rho) \Gamma(n + m + \rho +\half) \over 
                                                                     \Gamma(n + m + 1) \Gamma(n + 2m + 2\rho)^2}
\ena
where $x$ stands for $|z|^2$,  $\xi(x, n,m,\rho) = \dis \lambda^{2(n - m)}\, {2^{(2 n + 2 m + 2\rho)}\over 2 \sqrt{\pi}}  $  and 
$\omega_m = \dis {x^m g_m(x;a_r) \over \pi N_m^2(x; a_r)}$. 
After variable change $n + m \to s -1$ 
 \ and  using the Mellin inversion theorem in terms of Meijer's G-function (\ref{c15}), we deduce:
{\small
\bea \label{p22}
\omega_m(|z|^2;a_r) = {1\over 2\pi \sqrt{\pi}} {|z|^{2m} \over {{\cal N}_m(|z|^2; a_r)^2}}
 \lambda^{-2(1+2m)} 2 ^{2(1-\rho)}
G_{3,5}^{5,0}\left({|az|^2\over 4}\left|\ba{l}
 0,-1+2\rho + m,-1+2\rho + m \\ -m,-m,2\rho - 1, -1+ \rho,-1/2 + \rho   \ea \right.\right).
\ena}
The weight function (\ref{p22}) is positive for  the parameter $\rho > 0$ as shown  in Figure 1, where the curves are represented for  $\rho =  2$ and for $m = 0, 1, 2, 3$. All the functions are positive for $x = |z|^2\in \IR_+$ and tend asymptotically to the measure of the conventional CS $(m = 0)$. The measure has a singularity at $x =  0$  and tends to zero for $x \to \infty.$\\\\
 {\it (iv) Thermal statistics}\\ \\
 Consider the normalized density operator expression 
 \bea\label{dmatrix06}
 \rho^{(m)}   =  \frac{1}{Z}\sum_{n = 0} e^{-\beta E_n} |n+m\rangle \langle n+m|
\ena 
 in which the exponent $\beta E_n$ is re-cast    as follows:
 $\beta E_n =   \beta \lambda^2 \left[n^2+2n\rho \right]   = A n^2 - B_{\rho}n$
 where $ A = \beta \lambda^2$, $\, B_{\rho} = -2\beta \rho \lambda^2. $ Then, the energy exponential can  be expanded in the 
 power series,   (see for e.g., \cite{popov01}) such that
{\footnotesize 
\bea
 e^{-\beta E_n}=  e^{-A n}\left[\sum_{k= 0}^{\infty} \frac{(B_{\rho})^k}{k!} n^{2k}\right] &= & 
 \left\{\sum_{k= 0}^{\infty} \frac{(B_{\rho})^k}{k!} \left(\frac{d}{dA}\right)^{2k}\right\} \left(e^{-A}\right)^n 
=
\mbox{exp}\left[B_{\rho}\left(\frac{d}{dA}\right)^{2} \right]\left(e^{-A}\right)^n.
 \ena}
 Thereby, 
\bea\label{dmatrix07}
 \rho^{(m)}  &=& \frac{\mbox{exp}\left[B_{\rho}\left(\frac{d}{dA}\right)^{2} \right]}{Z}\sum_{n = 0}^{\infty} 
\left(e^{-A}\right)^n |n+m\rangle \langle n+m|.  
\ena
 From (\ref{p19}) and (\ref{dmatrix07}), we get, in terms 
of Meijer's G functions,
 the $Q$-distribution or Husimi 
distribution:
{\small
\bea 
 _{m}\langle z;m| \rho^{(m)} |z;m\rangle_m &=& \frac{\Gamma(2m+2\rho)}{\Gamma(m+\rho)\Gamma(m+\rho+1/2)}\frac{\Gamma(\frac{1}{2})}{2^{2(m + \rho - 1/2)}}\frac{\mbox{exp}\left[B_{\rho}\left(\frac{d}{dA}\right)^{2} \right]}{Z} \times \cr
&& \times \frac{ G_{3,5}^{1,3}\left(- \frac{\left(a|z|\right)^2}{4} e^{-A} \lv
 \ba{l}-m, 1 -2m-2\rho, 1 -2m-2\rho;   \\ 0 ; 0, 1 -m-\rho, 1 -m - 2\rho, 1/2-m-\rho  \ea   \right. \right)}
 { G_{3,5}^{5,0}\left(- \frac{\left(a|z|\right)^2}{4}\lv
 \ba{l}-m, 1-2m-2\rho, 1-2m-2\rho;   \\ 0;  0, 1 -m-\rho, 1-m-2\rho,  1/2 -m-\rho  \ea   \right. \right)}.
\ena}
The   angular integration achieved, taking    $x = |z|^2, $ the condition (\ref{thermal02}) supplies 
{\footnotesize
\bea 
\mbox{Tr}  \rho^{(m)} & = & \frac{\mbox{exp}\left[B_{\rho}\left(\frac{d}{dA}\right)^{2} \right]}{Z} \frac{2^{2[1-2\rho-m]}}{\lambda^{2(1+m)}} \int_{0}^{\infty} dx \, x^m 
G_{3,5}^{1,3}\left(- {\frac{|a|^2}{4} x e^{-A}} \lv
 \ba{l}-m, 1-2m-2\rho, 1-2m-2\rho;   \\ 0;  0, 1 - m - \rho, 1-m-2\rho,  1/2 -m-\rho  \ea   \right. \right)\times \cr
& \times& G_{3,5}^{5,0}\left(\frac{|a|^2}{4}x \lv
  \ba{l} \qquad ;0, -1+2m+\rho, -1 + 2\rho +m \\ -m, -m, 2\rho -1, -1 + \rho, -1/2 +\rho; \qquad  \ea \right. \right).
\ena}
Then, the  integral of Meijer's G-function product properties provides the partition function
\bea\label{part00}
 Z = \frac{2^{4(1-\rho)}}{(\lambda |a|)^{2(1+m)}}\mbox{exp}\left[B_{\rho}\left(\frac{d}{dA}\right)^{2} \right]\sum_{n = 0}^{\infty} 
\left(e^{-A}\right)^n.
\ena
From (\ref{thermal03}),   using the result 
$\langle n+m| \rho^{(m)} |n+m\rangle  =  {1\over Z}{\mbox{exp}\left[B_{\rho}\left(\frac{d}{dA}\right)^{2} \right]\left(e^{-A}\right)^n}$
and setting $\bar{n}_A =(e^{A} -1)^{-1}$, we get the following integration  equality
{\footnotesize
\bea 
&&\frac{1}{\bar{n}_A+1}\left(\frac{\bar{n}_A}{\bar{n}_A+1}\right)^n  \frac{\Gamma(n+1)^2\Gamma(n+m+2\rho) \Gamma(2n + 2m +2\rho)}
{\Gamma(n+m+1)\Gamma(n+2m+2\rho)^2} \frac{\sqrt{\pi}\lambda^{4(1+m)}}{2^{5-6\rho}|a|^{2(n-m-1)}}=   \int_{0}^{\infty} dx \, x^{n+m} P(x) \times \cr
&&  G_{3,5}^{5,0}\left(\frac{|a|^2}{4}x \lv
  \ba{l} \qquad ;0, -1+2m+\rho, -1 + 2\rho +m \\ -m, -m, 2\rho -1, -1 + \rho, -1/2 +\rho; \qquad  \ea \right. \right).\nonumber
\ena
}
After performing  the exponent change  $ n+m = s-1$ in order to get  the Stieltjes moment problem, 
we arrive at the $P$-function as 
{\small
\bea\label{dmatrix08}
& &P(|z|^2) =    \frac{1}{\bar{n}_A}\left(\frac{\bar{n}_A+1}{\bar{n}_A}\right)^m 
\frac{\lambda^2 |a|^{2(m+1)}}{2^{4(1-\rho)}} 
\frac{ G_{3,5}^{5,0}\left(\frac{\bar{n}_A+1}{\bar{n}_A}\frac{|az|^2}{4} \lv
  \ba{l} \qquad ;0, -1+2\rho + m, -1 + 2\rho + m \\ -m, -m, 2\rho -1, -1 + \rho, -1/2 +\rho; \qquad  \ea \right. \right)}
{ G_{3,5}^{5,0}\left(\frac{|az|^2}{4} \lv
  \ba{l} \qquad ;0, -1+2\rho + m, -1 + 2\rho + m \\ -m, -m, 2\rho -1, -1 + \rho, -1/2 +\rho; \qquad  \ea \right. \right)}
\ena
} 
which  obeys the normalization to unity condition (\ref{thermal04}).\\\\
Then, the diagonal representation of the normalized density operator in terms of the  PA-SIPCS
projector (\ref{thermal03}) takes the form
 \bea\label{dmatrix09} 
 \rho^{(m)}  = \frac{1}{\bar{n}_A}\left(\frac{\bar{n}_A+1}{\bar{n}_A}\right)^m
 \frac{\lambda^2 |a|^{2(m+1)}}{2^{4(1-\rho)}} \int_{\IC} d^2 z \, \omega_m(|z|^2; a_r) 
|z;a_r\rangle_m \mathfrak S_{3,5}^{5,0}(|z|^2; {\bar{n}_A}) \, _{m}\langle z;a_r|     
\ena
with $ \mathfrak S_{3,5}^{5,0}(|z|^2, {\bar{n}_A})$ -  the Meijer's G-functions quotient given in (\ref{dmatrix08}).
\ni Using  the relations (\ref{dmatrix08}), (\ref{dmatrix09}), and the definition (\ref{obaverage00}),   the 
pseudo-thermal expectation values of the  operator $N$ and  its 
  square  are given by
  \bea
\langle   N\rangle^{(m)} &=& \left(\frac{|a|^2}{4}\right)^{m+1}\frac{1}{\lambda^{2(m-1)}}\frac{m(m+2\rho)}{(m+1)(m+1+2\rho)}\cr
&& \times \left[1 + \left(\frac{1}{m+1}+\frac{1}{m+1+2\rho}\right)\bar{n} + \frac{1}{(m+1)(m+1+2\rho)}\left(\frac{\bar{n}}{1-e^{-\beta}}+ \bar{n}^2\right)\right]
\ena  
   \bea
&& \langle N^2 \rangle^{(m)}  = \left(\frac{|a|^2}{4}\right)^{m+1}\frac{1}{\lambda^{2(m-2)}}\left[\frac{m(m+2\rho)}{(m+1)(m+1+2\rho)}\right]^2 \times \left\{
1 + 2\left(\frac{1}{m+1}+\frac{1}{m+1+2\rho}\right)\bar{n} + \right.\cr
& & \left. +\left(\frac{1}{(m+1)^2} + \frac{1}{(m+1+2\rho)^2} + \frac{4}{(m+1)(m+1+2\rho)}\right) 
\left(\frac{\bar{n}}{1-e^{-\beta}} + \bar{n}^2\right)+ \right. \cr
& & 
\left.  + 2\left(\frac{1}{(m+1)^2(m+1+2\rho)} + \frac{1}{(m+1)(m+1+2\rho)^2}\right)
\left[\frac{\bar{n}}{(1-e^{-\beta})^2} + 
\frac{4\bar{n}^2}{1-e^{-\beta}} + \bar{n}^3\right] + \right.\cr
& &   + \frac{1}{(m+1)^2(m+1+2\rho)^2}
\left. \left[\frac{\bar{n}}{(1-e^{-\beta})^3} + \frac{11\bar{n}^2}{(1-e^{-\beta})^2} + \frac{11\bar{n}^3}{1-e^{-\beta}} + \bar{n}^4\right]\right\}
 \ena 
 where $\bar{n} = (e^{-\beta} -1)^{-1}. $ Thereby, 
  \bea
(g^2)^{(m)}  &=&  1+ \left\{\left(\frac{1}{(m+1)} + 
 \frac{1}{(m+1+2\rho)}\right)^2 \frac{\bar{n}}{(1-e^{-\beta})}+ \right. \cr &&  \left.
\left(\frac{1}{(m+1)^2(m+1+2\rho)} + \frac{1}{(m+1)(m+1+2\rho)^2}\right)
\left[\frac{2\bar{n}}{(1-e^{-\beta})^2} + \frac{6\bar{n}^2}{1-e^{-\beta}}\right] + \right. \cr &&  \left. 
\frac{1}{(m+1)^2(m+1+2\rho)^2}\left[\frac{\bar{n}}{(1-e^{-\beta})^3} + \frac{10\bar{n}^2}{(1-e^{-\beta})^2} + \frac{9\bar{n}^3}{1-e^{-\beta}}\right]\right\} \cr
&&\times \frac{1}{\left(\frac{|a|^2}{4}\right)^{-(m+1)} \lambda^{2m}\left(\langle   N \rangle^{(m)}\right)^2} - \frac{1}{\left(\frac{|a|^2}{4}\right)^{-\frac{m +1}{2}}\lambda^m \langle N \rangle^{(m)}}.
\ena
Then, the   thermal analogue of the Mandel parameter is  given by
\bea
 Q^{(m)} & = &  \left(\frac{|a|^2}{4}\right)^{-\frac{m+1}{2}}\lambda^m\langle  N \rangle^{(m)}\left[(g^2)^{(m)}-1\right] \cr
 &=&  \left\{\left(\frac{1}{(m+1)} +  \frac{1}{(m+1+2\rho)}\right)^2 \frac{\bar{n}}{(1-e^{-\beta})}+ \right. \cr && \left.
\left(\frac{1}{(m+1)^2(m+1+2\rho)} + \frac{1}{(m+1)(m+1+2\rho)^2}\right) 
\left[\frac{2\bar{n}}{(1-e^{-\beta})^2} + \frac{6\bar{n}^2}{1-e^{-\beta}}\right] + \right. \cr && \left.
 \frac{1}{(m+1)^2(m+1+2\rho)^2}\left[\frac{\bar{n}}{(1-e^{-\beta})^3} + \frac{10\bar{n}^2}{(1-e^{-\beta})^2} + \frac{9\bar{n}^3}{1-e^{-\beta}}\right]\right\} \cr
&&\times \frac{1}{\left(\frac{|a|^2}{4}\right)^{-\frac{m+1}{2}}\lambda^m \langle N \rangle^{(m)}} - 1.
\ena
\subsection{Second choice  of the functional ${\cal Z}_j$}
\ni We  now take  
${\cal Z}_j = \sqrt{g(a_1; \kappa, \kappa)g(a_1; \kappa, 0) } \, e^{- i\alpha {\cal R}(a_1)}$
with $\kappa$ a real constant and where we use the auxiliary function \cite{Alex}
$g(a_j ; c, d) = c a_j + d$,  $c$ and $d$ being real constants. From the potential parameter relations (\ref{p11}) we obtain:  
\bea \label{p28}
\prod_{k = m}^{n+m-1} g(a_{j + k};c,d)  = {2c}^n 
 {\Gamma(n +  m + {a_1\over 2} + j - 1 + d/2c) \over \Gamma(m + {a_1\over 2} + j - 1 + d/2c) }.
\ena
Setting $a_1 = 2\rho$, we have:
\bea \label{p29}
\prod_{k = m}^{n+m-1} {\cal Z}_{j + k} = \left[\kappa^{2n}{\Gamma(2n + 2 m + 2\rho) \over \Gamma(2 m + 2\rho)} \right]^{\half}e^{-i \alpha E_n} 
\ena 
with the eigen-energy $E_n$ given by (\ref{p15}). 
Inserting Eqs. (\ref{p29}), (\ref{p12}) and (\ref{p13}) in the expansion coefficient (\ref{c5}), we obtain  
\bea \label{p30}
K_n^m(a_r) = \left[ {1\over \kappa^{2m}}{ \Gamma(n + 1)^2\, \Gamma(n + m + \nu + 1)\Gamma( 2 m + \nu + 1) \over \Gamma(n + m + 1)\Gamma(n +2 m +\nu + 1)^2 
 }\right]^{\half}e^{i \alpha E_n}, 
\ena
where we assume $\lambda = {1\over a} =  \kappa$ and $\rho = {\nu\over 2} +  \half, \nu\ge  1 $ in (\ref{p12}) and (\ref{p13}). 
For $m = 0$, we recover the coefficient $h_n$ in \cite{Alex}:
\bea \label{p31}
K_n^0(a_r) = \left[{\Gamma(n+1) \Gamma(\nu +1) \over\Gamma(n +\nu+1)}\right]^{\half} e^{i \alpha E_n} = h_n(a_r)\pt
\ena
 {\it (i) Normalization} \\\\
 The normalization factor in terms of hypergeometric and Meijer's G-functions is 
{\small
\bea \label{p32}
 {\cal N}_m(|z|^2; a_r) & = & 
\kappa^{2m} \Gamma(m+1){\Gamma(2m+ \nu+1) \over \Gamma(m + \nu +1)} 
 \left[{}_3F_2\left(\ba{lcr}m+1, 2m+\nu +1, 2m+\nu +1 &;&\\
 1, m+\nu +1& ;& |z|^2\ea\right)\right]^{-\half} \\ \label{p33}
 {\cal N}_m(|z|^2; a_r)& = &
\left[{ \kappa^{2m} \over \Gamma(2m+\nu + 1)} G_{3,3}^{1,3}\left(-|z|^2\left|
\ba{ccc}-m, -2m-\nu, -2m-\nu   \quad  \\ 0 , 0, - m - \nu \ea \right.\right)\right]^{-\half}.
\ena}
The explicit form of the PA-SIPCS,  defined for $|z| < 1$,  is provided by: 
{
\bea \label{p34}
 \lv z; a_r\rs_m = {\cal N}_m (|z|^2; a_r) \sum_{n = 0}^\infty 
\sqrt{ \kappa^{2m}{\Gamma(n + m + 1)\Gamma(n +2 m +\nu + 1)^2 \over 
\Gamma(n + 1)^2 \Gamma(n + m +\nu + 1)  \Gamma ( 2 m + \nu + 1) }}z^{n}e^{-i \alpha E_n} \lv n + m\rs.
\ena}
For $m=0$, we recover  the normalization factor 
\bea \label{p35}
{\cal N}_0(|z|^2; a_r) = {}_1F_0(\nu + 1;-; |z|^2)^{-\half}  = (1 - |z|^2)^{-1/2 - \nu/2 } =
 {\cal N}(|z|^2; a_r)
\ena
obtained in \cite{Alex}.  For $m = 0$, the PA-SIPCS is  reduced  to the  SIPCS  
\bea \label{p36}
\lv z; a_r\rs = (1 - |z|^2)^{(\nu +1)/2}\,
 \sum_{n=0}^\infty \sqrt{\Gamma(n + \nu + 1) \over \Gamma(\nu + 1) \Gamma(n + 1)}\, 
e^{-i\alpha E_n}\, \lv \Psi_n\rs
\ena
obtained in \cite{Alex} and in \cite{Kinani} as CS of Klauder-Perelomov's type  for the PT-I.\\
 {\it (ii) Non-orthogonality} \\\\
 The inner product of two different PA-SIPCS $\lv z; a_r\rs_m$ and $\lv z';a_r\rs_{m'}$
 is given by:
{
\beano
 _{m'}\li z';a_r\rv \left. z; a_r\rs_m & = & \chi(z',z, m, m',\nu)\, 
_3F_2\left(\ba{lcr}m +1, m+ m' +\nu+1, 2m+\nu+1 &; &\\
m - m'+ 1, m+\nu+1 &; &{ z'}^\star z \ea\right)
\enano}
where 
{
\beano
 \chi(z',z, m, m',\nu') & = &{\cal N}_{m'}(|z'|^2;a_r)\ {\cal N}_m(|z|^2;a_r) 
\ { {{ z'}^\star}^{(m-m')} \kappa^{(m+m')} \over \sqrt{\Gamma(2m +\nu+1) \Gamma(2m'+\nu +1) }}\times \\
 &  & \times 
 {\Gamma(m +1)  \Gamma(m +m'+\nu+1)  \Gamma(2m +\nu+1) \over \Gamma(m -m'+1)  \Gamma(m +\nu+1)}
\ e^ {i\alpha(E_n-E_{n+m-m'})}.\\
\enano }
{\it (iii) Overcompleteness}\\\\
\ni Following the steps  of section 2.3,  we obtain the weight-function of the  PA-SIPCS (\ref{p34})  as 
{
{\small
\bea \label{p36}
\omega_m(|z|^2; a_r) & = &{1 \over \pi} G_{3,3}^{1,3}\left(-|z|^2\left|
\ba{l}-m, -2m-\nu, -2m-\nu   \quad  \\ 0 , 0, - m - \nu \ea \right.\right)\,
G_{3,3}^{3,0}\left(|z|^2\left|
\ba{l} m, 2 m + \nu, 2 m + \nu  \\0, 0, m+ \nu   \ea \right.\right) .
\ena}
We recover, for $m = 0$, the result:
\bea \label{p37}
\omega_0(|z|^2; a_r) & = & {\Gamma(\nu + 1) \over \pi}\, 
_1F_0(\nu + 1;-; |z|^2)\, 
G_{1,1}^{1,0}\left(|z|^2\left|
\ba{ccc} \quad & ; & \nu  \\0  & ; &   \ea \right.\right) = {\nu \over \pi} (1 - |z|^2)^{-2}
\ena
obtained in \cite{Alex} for the corresponding ordinary SIPCS. \\\\ 
{\it (iv) Thermal statistics}\\ \\
Since the eigen-energy $E_n$ (\ref{p15}) is the same as previously, we start by maintaining the relations (\ref{dmatrix06})-(\ref{dmatrix07}). 
 From (\ref{p34}) and (\ref{dmatrix07}), we get, in terms 
 of Meijer's G functions,
 the $Q$-distribution or Husimi 
distribution
{
\bea 
 && _{m}\langle z;m| \rho^{(m)} |z;m\rangle_m = \frac{\mbox{exp}\left[B_{\rho}\left(\frac{d}{dA}\right)^{2} \right]}{Z}
\frac{G_{3,3}^{1,3}\left(-|z|^2e^{-A}\lv
 \ba{l}-m,-2m-\nu, -2m-\nu;   \\ 0 ; 0, -m-\nu   \ea   \right. \right)}{G_{3,3}^{1,3}\left(-|z|^2\lv
 \ba{l}-m,-2m-\nu, -2m-\nu;   \\ 0 ; 0, -m-\nu   \ea   \right. \right)}.
\ena}
The   angular integration achieved, taking    $x = |z|^2, $ the condition (\ref{thermal02}) supplies 
{
\bea 
\mbox{Tr}  \rho^{(m)} & = & \frac{\mbox{exp}\left[B_{\rho}\left(\frac{d}{dA}\right)^{2} \right]}{Z}  \int_{0}^{\infty} dx \, x^m 
G_{3,3}^{3,0}\left(|z|^2 \lv
 \ba{l}\qquad; m,2m+\nu, 2m+\nu  \\ 0,  0, m+\nu; \qquad   \ea   \right. \right)\times \cr
& & G_{3,3}^{1,3}\left(-|z|^2 e^{-A}\lv
 \ba{l}-m,-2m-\nu, -2m-\nu;   \\ 0 ; 0, -m-\nu   \ea   \right. \right).
\ena}
Then, the  integral of Meijer's G-function product properties provides the partition function expression
$ Z = \mbox{exp}\left[B_{\rho}\left(\frac{d}{dA}\right)^{2} \right]\dis\sum_{n = 0}^{\infty} 
\left(e^{-A}\right)^n.$
From (\ref{thermal03}), taking $\bar{n}_A =(e^{A} -1)^{-1}$, we get the following integration  equality
{\small
\bea 
&&\frac{1}{\bar{n}_A+1}\left(\frac{\bar{n}_A}{\bar{n}_A+1}\right)^n  \frac{\Gamma(n+1)^2\Gamma(n+m+\nu+1)}
{\Gamma(n+m+1)\Gamma(n+2m+\nu+1)^2} = \int_{0}^{\infty} dx \, x^{n} P(x) 
 G_{3,3}^{3,0}\left(|z|^2 \lv
 \ba{l}\qquad; m,2m+\nu, 2m+\nu  \\ 0,  0, m+\nu; \qquad   \ea   \right. \right).\nonumber
\ena
}
Finally, we arrive at the $P$-function: 
{
\bea\label{dmatrix122}
& &P(|z|^2) =    \frac{1}{\bar{n}_A} 
\frac{G_{3,3}^{3,0}\left(\frac{\bar{n}_A+1}{\bar{n}_A}|z|^2 \lv
 \ba{l}\qquad; m,2m+\nu, 2m+\nu  \\ 0,  0, m+\nu; \qquad   \ea   \right. \right)}
{G_{3,3}^{3,0}\left(|z|^2 \lv
 \ba{l}\qquad; m,2m+\nu, 2m+\nu  \\ 0,  0, m+\nu; \qquad   \ea   \right. \right)}
\ena
} 
which  obeys the normalization to unity condition (\ref{thermal04}).\\\\
Then, the diagonal representation of the normalized density operator in terms of the  PA-SIPCS
projector (\ref{thermal03}) takes the form
 \bea\label{dmatrix124}
 \rho^{(m)}  = \frac{1}{\bar{n}_A} \int_{\IC} d^2 z \, \omega_m(|z|^2; a_r) 
|z;a_r\rangle_m \mathfrak S_{3,3}^{3,0}(|z|^2; {\bar{n}_A}) \, _{m}\langle z;a_r|     
\ena
with $ \mathfrak S_{3,3}^{3,0}(|z|^2, {\bar{n}_A})$  - the Meijer's G-functions quotient given in (\ref{dmatrix122}). \ni Using  the relations (\ref{dmatrix122}), (\ref{dmatrix124}), and the definition (\ref{obaverage00}),   the 
pseudo-thermal expectation values of the  operator $N$ and  its 
  square  are given by
  \bea
\langle   N \rangle^{(m)} = \kappa^2 m (m+\nu+1)\left[1 + \left(\frac{1}{m+1}+\frac{1}{m+\nu+2}\right)\bar{n} + \frac{1}{(m+1)(m+\nu+2)}\left(\frac{\bar{n}}{1-e^{-\beta}}+ \bar{n}^2\right)\right]
\ena  
  \bea
 \langle N^2 \rangle^{(m)}  &=& \kappa^4 m^2 (m+\nu+1)^2
\left\{1 + 2\left(\frac{1}{m+1}+\frac{1}{m+\nu+2}\right)\bar{n} + \right. \cr
 && \left. \left(\frac{1}{(m+1)^2} + \frac{1}{(m+\nu+2)^2}   + \frac{4}{(m+1)(m+\nu+2)}\right) \left(\frac{\bar{n}}{1-e^{-\beta}}+ \bar{n}^2\right) + \right. \cr
& & \left. 2\left(\frac{1}{(m+1)^2(m+\nu+2)} + \frac{1}{(m+1)(m+\nu+2)^2}\right)
\left(\frac{\bar{n}}{(1-e^{-\beta})^2} + \frac{4\bar{n}^2}{1-e^{-\beta}} + \bar{n}^3\right) + \right.\cr
 && \left. \frac{1}{(m+1)^2(m+\nu+2)^2}\left[\frac{\bar{n}}{(1-e^{-\beta})^3} + \frac{11\bar{n}^2}{(1-e^{-\beta})^2} + \frac{11\bar{n}^3}{1-e^{-\beta}} + \bar{n}^4\right]\right\}.
 \ena 
 Thereby, 
  \bea
(g^2)^{(m)}  &=&  1+ \left\{\left(\frac{1}{(m+1)} +  \frac{1}{(m+\nu+2)}\right)^2 \frac{\bar{n}}{(1-e^{-\beta})} + \left(\frac{1}{(m+1)^2(m+\nu+2)}\right. \right. \cr
&& \times \left. \left. + \frac{1}{(m+1)(m+\nu+2)^2}\right)
\left[\frac{2\bar{n}}{(1-e^{-\beta})^2} + \frac{6\bar{n}^2}{1-e^{-\beta}}\right] + \frac{1}{(m+1)^2(m+\nu+2)^2} \right.\cr
&&\left. \left[\frac{\bar{n}}{(1-e^{-\beta})^3} + \frac{10\bar{n}^2}{(1-e^{-\beta})^2} + \frac{9\bar{n}^3}{1-e^{-\beta}}\right]\right\} \frac{1}{\left(\langle   N \rangle^{(m)}\right)^2} - \frac{1}{\langle N \rangle^{(m)}}.
\ena
Then, the   thermal analogue of the Mandel parameter is given by
\bea
 Q^{(m)}  &=&  \langle  N\rangle^{(m)}\left[(g^2)^{(m)}-1\right] = 
 \left\{\left(\frac{1}{(m+1)} +  \frac{1}{(m+\nu+2)}\right)^2 \frac{\bar{n}}{(1-e^{-\beta})}+ \right.\cr
& &\left.  \left(\frac{1}{(m+1)^2(m+\nu+2)} + \frac{1}{(m+1)(m+\nu+2)^2}\right)
\left(\frac{2\bar{n}}{(1-e^{-\beta})^2} + \frac{6\bar{n}^2}{1-e^{-\beta}}\right) \right. + \cr
& & \left.  \frac{1}{(m+1)^2(m+\nu+2)^2}\left[\frac{\bar{n}}{(1-e^{-\beta})^3} + \frac{10\bar{n}^2}{(1-e^{-\beta})^2} + \frac{9\bar{n}^3}{1-e^{-\beta}}\right]\right\}
\frac{1}{\langle   N \rangle^{(m)}} - 1.
\ena
\section{Concluding remarks}

In this contribution paper, we have shown the use of the  shape invariant potential  method to construct generalized coherent states  for photon-added  particle systems under P\"oschl-Teller potentials.
These states  have been  fully characterized and discussed from both mathematics  
and physics points of view. This algebro-operator method can be exploited to investigate a larger class of solvable potentials.



\begin{thebibliography}{99} 


\bibitem{Aga}{G. S. Agarwal and K. Tara, Nonclassical properties of states generated by the excitations on a coherent state,   {\it phys. Rev A. }\, {\bf 43}, 492 (1991) ; \\
G. S. Agarwal and K. Tara,  Nonclassical character of states exhibiting no squeezing or sub-Poissonian statistics, 
 {\it phys. Rev A. }\, {\bf 46}, 485 (1992)}
     \bibitem{Alex}{A. N. F. Aleixo and A. B. Balantekin,
     {An algebraic  construction of generalized    coherent states for shape-invariant potentials},
     {\it J. Phys. A: Math. Gen.}\,  {\bf 37}, 8513 (2004)}
\bibitem{Antoine}{J.-P. Antoine, J.-P. Gazeau, P.M. Monceau, J. R. Klauder and K. A. Penson, 
Temporally stable coherent states for infinite well
and P\"oschl-Teller potentials, 
  {\it J. Math. Phys.}  {\bf 42},  2349 (2001)}
\bibitem{Ali95}{S. T. Ali, J.-P. Antoine, J.-P. Gazeau U. A.  and Mueller,   
Coherent states and their generalizations : a mathematical overview, 
{\it Rev. Math. Phys.},  {\bf 7},  1013 (1995)}
\bibitem{Aremua1}{I. Aremua, J.-P. Gazeau and M. N. Hounkonnou, 
Coherent states for Landau Levels : algebraic and thermodynamical properties
 {\it J. Phys. A: Math. Gen.}  {\bf 45},  335302 (2012)}
\bibitem{Aremua2}{I. Aremua, M. N.  Hounkonnou and E. Balo\"{\i}tcha,  
Supersymmetric vector coherent states for systems with Zeeman coupling and spin-orbit interactions, 
{\it Rep. Math. Phys.}  {\bf 45}(2),  247 (2015)}
\bibitem{Bal}{A. B. Balantekin,
 Algebraic approach to shape invariance,
      {\it  Phys. Rev. A} {\bf 57} (6), 4188 (1998); \\
      A. B. Balantekin, M. A. C\^andido Ribeiro and A. N. F. Aleixo,
    Algebraic nature of shape-invariant and self-similar potentials,
      {\it J. Phys. A: Math. Gen.}  {\bf 32}, 2785 (1999); \\
      E. D. Filho and  M. A. C\^andido Ribeiro, 
Generalized Ladder Operators for Shape-invariant Potentials,
{\it Phys. Scr. }, {\bf 64} (6),  548 (2001)}
\bibitem{Ban}{M. Ban, 
Photon statistics of conditional output states of lossless beam splitter,
 {\it J. Mod. Opt.}\, {\bf 43}, 1281 (1996)}
\bibitem{Bergeron}{H. Bergeron, P. Siegl and A. Youssouf, 
 New SUSYQM coherent states for P\"oschl-Teller potentials: a detailed mathematical analysis, 
{\it J. Phys. A: Math. Theor.}, {\bf  45}, 244028 (2012).}
\bibitem{Cooper}{F. Cooper, A.  Khare  and  U. Sukhatme,
  Supersymmetry and   quantum mechanics,
{\it  Phys. Rep.}  {\bf 251},  267 (1995)}
\bibitem{Dab}{J. W. Dabrowska, A. Khare and U. P. Sukhatme,
  Explicit wavefunctions for
       shape-invariant  potentials by operator techniques,
{\it  J. Phys. A: Math. Gen.} {\bf 21}, L195 (1988)}
\bibitem{Dakna}{M. Dakna, T. Anhut, T. Opatrny, L. Kn\"oll D.-G. and Welsch D-G,
Generating Schr\"odinger-cat-like states by means of conditional measurements on a beam splitter
  {\it Phys. Rev.A.} {\bf 55}, 3184 (1997) }\\
{M. Dakna, L. Kn\"oll   and D.-G. Welsch, 
 Photon-added state preparation via conditional measurement on abeam splitter, 
 {\it Opt. Commun.} {\bf 145}, 309 (1998)}
\bibitem{Daskalo}{C. Daskaloyannis,
Generalized deformed oscillator 
corresponding to the modified P\"oschl-Teller energy-spectrum.
 {\it J. Phys. A: Math. Gen.} {\bf 25}, 2261. (1992) }
 \bibitem{Dutt5}{R. Dutt, A.  Khare  and  U. P. Sukhatme, 
Exactness of Supersymmetric WKB spectrum for shape invariant potentials, 
  {\it  Phys. Lett.}  {\bf B 181}, 295 (1986)} 
 \bibitem{Fukui} T. Fukui and N.  Aizawa, 
Shape-invariant potentials and associated  coherent states, 
{\it Phys. Lett. A}\, {\bf 180}, 308 (1993)
\bibitem{Gend}{L. E. Gendenshtein,
Derivation of exact spectra of the schr\"odinger equation by means of supersymmetry,
 {\it JETP Lett.} {\bf 38}, 356 (1983)}
\bibitem{Glauber}{R. J. Glauber, The quantum theory of optical coherence,
 {\it Phys. Rev.} {\bf 130}, 2529 (1963); \\
R. J. Glauber,
Coherent and Incoherent States of the Radiation Field, 
 {\it Phys. Rev.} {\bf 131}, 2766  (1963)}
\bibitem{Sama}{M. N. Hounkonnou, S. Arjika  and E. Balo\"{\i}tcha,  
P\"oschl-Teller Hamiltonian: Gazeau-Klauder type coherent states, related statistics, and geometry, 
{\it J. Math. Phys.} {\bf 55}, 123502 (2014)}
\bibitem{HSA}{M. N. Hounkonnou, K. Sodoga and E. Azatassou, 
Factorization of Sturm-Liouville operators: solvable potentials and underlying algebraic structure,  
{\it J. Phys. A: Math. Gen.} {\bf 38}, 371 (2005)}
\bibitem{Sodoga}{M. N. Hounkonnou  and  K. Sodoga, 
 Generalized coherent states for associated hypergeometric-type functions, 
 {\it J. Phys. A: Math. Gen.}  {\bf 38},  7851 (2005)}
\bibitem{Infeld}{L. Infeld and  T. E. Hull, The factorization method,  {\it Rev. Mod. Phys.} {\bf 23},  28 (1951)}
\bibitem{Kinani}{A. H. El Kinani and  M. Daoud, 
Coherent states \`a la klauder-Perelomov for the P\"osch-Teller potentials, 
 {\it Phys. Lett. A. }\, {\bf 283}\, 291 (2001);\\ 
M. Daoud,  Photon-added coherent states for exactly solvable Hamiltonians, 
 {\it Phys. Lett. A. }\, {\bf 305}\, 135 (2002)}
\bibitem{Suk}{A. Khare and U. P. Sukhatme,
   { New shape invariant potentials in supersymmetry quantum mechanics},
     {\it J. Phys. A: Math. Gen.}  {\bf 26}, L901 (1991)}  
 \bibitem{6derniers}{J. R. Klauder, K. A. Penson and J.-M. Sixderniers, 
Constructing coherent states through solutions of Stieltjes and Hausdorff moment problems
{\it Phys. Rev. A}\, {\bf 64}\, 013817 (2001)}
\bibitem{Mandel95}{L. Mandel and E. Wolf,  {\it Optical coherence and quantum optics} (Cambridge University Press, Cambridge, 1995)}
\bibitem{Mathai}{A. M. Mathai and R. K.  Saxena,  {\it Generalized Hypergeometric Functions with Applications in Statistics and
Physical Sciences} (Lecture Notes in Mathematics vol 348) (Springer, Berlin, 1973)}
\bibitem{Penson99}{K. A. Penson  and A. I. Solomon 
New generalized coherent states, {\it J. Math. Phys.}\, {\bf 40}, 2354 (1999)}
\bibitem{Pere}{A. M. Perelomov A M\, {\it Generalized coherent states and their applications}  (Springer, Berlin, 1986)}
 \bibitem{Popov}{D. Popov,  
Photon-added Barur-Girardello coherent states of the pseudoharmonic oscillator, 
   {\it J. Phys Phys. Math. Gen}\,  {\bf 35} \,  7205 (2002)}
\bibitem{popov01} {D. Popov, I. Zaharie and S. H. Dong,  
Photon-added coherent states for the Morse oscillator
 {\it Czech. J. Phys.}\,  {\bf 56},  157} (2006); \\ 
 {D. Popov,  
Some Properties of Generalized Hypergeometric Thermal Coherent States, 
{\it Electron. J. Theor. Phys. (EJTP)}\,  {\bf 3}(11),  123 (2006)}
\bibitem{Poschl-Teller}{G. P\"oschl, E. Teller, 
``Bemerkungen zur Quantenmechanik des Anharmonischen Oszillators'', 
{\it Z. Phys.} {\bf 83}, 143 (1933)}
\bibitem{Rosen}{N. Rosen, Ph.M. Morse,
On the vibrations of polyatomic molecules, 
 {\it Phys. Rev.} {\bf 42},   210, (1932)}
\bibitem{Scarf}{F.L. Scarf,
New Soluble Energy Band Problem
 {\it Phys. Rev.} {\bf 112}, 1137 (1958) }
\bibitem{Schrodinger}{E. Schr\"odinger, 
 The continuous transition from micro- to macro-mechanics, 
  {\it Naturwiss}\, {\bf 14}, 664 (1926)}
\bibitem{Penson}{J.-M. Sixderniers and K. A. Penson K A,  
On the completeness of photon-added coherent states, 
  {\it J. Phys. A. Math. Gen}\,  {\bf 34} \,  2859 (2001)}
\bibitem{SAH_prep}{K. Sodoga, M. N. Hounkonnou and Aremua I,
{Photon-added coherent states for shape invariant systems}, 
 arxiv: 703.01629v1, (2017)}
\bibitem{Teschl}{G. Teschl, Mathematical methods in quantum mechanics:  with application to Schr\"odinger operators, Graduate Studies in Mathematics, Vol 157, 2nd edition,   American Mathematical Society
(1999)}

\end{thebibliography}
\end{document}